\documentclass[trackchanges]{aastex701}

\begin{document}

\title{A Sample of Long-Period Binaries with Dormant Compact-Object Candidates from LAMOST DR13 MRS and \textit{Gaia} DR3}

\author[orcid=0009-0006-1081-8203]{Junxing Zhou}
\affiliation{Department of Astronomy, Xiamen University, Xiamen, Fujian 361005, People’s Republic of China}
\email{jxzhou@stu.xmu.edu.cn}

\author[orcid=0000-0003-3137-1851]{Wei-Min Gu} 
\affiliation{Department of Astronomy, Xiamen University, Xiamen, Fujian 361005, People’s Republic of China}
\email[show]{guwm@xmu.edu.cn}


\begin{abstract}


Identifying dormant compact objects in non-interacting binaries is crucial for constraining the hidden population of stellar remnants. We present a systematic search for long-period compact-object candidates by combining \textit{Gaia} DR3 astrometry with multi-epoch LAMOST DR13 medium-resolution spectroscopy. Starting from sources with elevated RUWE values, we required at least five LAMOST spectroscopic epochs and peak-to-peak RV variations exceeding $20\,{\rm km\,s^{-1}}$. After RV extraction, orbital fitting, and subsequent screening, we obtained Bayesian Keplerian orbital solutions for 19 long-period single-lined binary candidates, with periods spanning approximately 30--1500 days. Multi-wavelength spectral energy distribution fitting was used to constrain the visible stars and evaluate the expected luminosities of unseen companions under a main-sequence assumption. Two systems are identified as the highest-priority compact-object candidates in our sample, each with a minimum companion mass exceeding $1\,M_{\odot}$. Their spectra and optical-to-infrared photometry show no detectable secondary-light contribution, although such a contribution would be expected if the unseen companions were normal main-sequence stars. These systems therefore represent the strongest compact-object candidates in our sample, although further observations are required to distinguish massive white dwarfs or neutron stars from faint stellar companions or unresolved hierarchical systems. We also identify five intermediate-priority and 12 lower-priority systems for follow-up spectroscopy, ultraviolet imaging, and future astrometric constraints.

\end{abstract}



\section{INTRODUCTION}
\label{sec:intro}

Theoretical binary population synthesis models and cosmological simulations predict a vast, hidden population of compact object binaries within the Milky Way, including over a million binary black holes \citep{2002ApJ...572..407B, 2018MNRAS.480.2704L}. However, the known sample of these extreme compact objects remains frustratingly sparse. Historically, the discovery of stellar-mass remnants has been heavily biased toward interacting binary systems, such as X-ray binaries, where the compact object is actively accreting matter from a companion \citep{2006ARA&A..44...49R, 2016non_gaia}. Yet, the vast majority of compact objects in binary systems are expected to exist in wide, detached, and non-interacting configurations. In the absence of luminous accretion disks or significant high-energy emission, this vast population of dormant compact objects remains largely invisible to traditional X-ray and radio surveys, creating a critical gap in our understanding of compact object formation and binary evolution.

Unmasking this hidden population requires shifting our observational paradigm from accretion-driven signatures to purely kinematic and dynamical tracers. In an unresolved binary system, a massive, dark companion will induce a gravitational wobble on its luminous, visible primary star. The \textit{Gaia} mission has revolutionized this approach. Specifically, when a standard single-star astrometric solution is forced upon a binary system, the unresolved orbital motion introduces significant astrometric perturbations \citep{2023A&A...674A...1G}. This deviation is effectively parameterized by the Renormalized Unit Weight Error (RUWE) in the \textit{Gaia} Data Release 3 (DR3) archive. While a well-behaved single star typically exhibits a $\mathrm{RUWE} \approx 1.0$, an elevated RUWE can indicate unresolved multiplicity or astrometric perturbations, although it may also arise from crowding, calibration residuals, or other astrometric systematics \citep{RUWE2020, RUWE2022}. We therefore use RUWE as an efficient pre-selection indicator rather than as direct evidence for a compact companion. 

In this work, we present a systematic search for long-period single-lined binaries that may host compact objects by combining \textit{Gaia} astrometry with large-scale spectroscopic monitoring. We use the \textit{Gaia} DR3 RUWE as an efficient initial selector and then constrain the line-of-sight orbital dynamics using time-domain LAMOST MRS spectra. This combination mitigates the selection biases of interaction-driven surveys and allows us to identify 19 long-period binary candidates for follow-up studies.

\section{DATA AND ANALYSIS} \label{sec:data}

\subsection{Astrometric Initial Sample\label{subsec:astrometric}}

In an unresolved binary system hosting a dormant compact object, the orbital motion of the visible primary around the system's barycenter introduces significant astrometric perturbations. Consequently, the standard five-parameter single-star astrometric solution applied by \textit{Gaia} DR3 \citep{2023A&A...674A...1G} often yields a poor fit for such systems. This deviation is effectively parameterized by the RUWE. While a $\mathrm{RUWE} \approx 1.0$ indicates a well-behaved single star, an elevated RUWE may indicate unresolved multiplicity or astrometric perturbations. RUWE is preferentially sensitive to longer-period unresolved binaries, particularly those with orbital periods comparable to the Gaia observing baseline \citep{2020MNRAS.496.1922B,2022MNRAS.513.2437P, ruwe2024A&A...688A...1C}. Therefore, combining radial-velocity and astrometric information provides complementary sensitivity across different orbital-period regimes and helps reduce period-dependent selection biases.

Although $\mathrm{RUWE} > 1.25$ has been suggested as the nominal threshold for significant astrometric excess in Gaia DR3 \citep{2026MNRAS.548ag654G}, we adopt the more restrictive cut of $\mathrm{RUWE} > 1.4$ to construct a conservative sample of sources with pronounced astrometric deviations. This choice is intended to improve the purity of the selected sample at the expense of some completeness. This criterion provides an efficient way to enrich the sample in unresolved binaries, but it does not by itself uniquely identify the origin of the astrometric excess. This selection yielded a parent sample of 64,502,228 sources.

\subsection{Spectroscopic Data Processing \label{subsec:spectroscopy}}

To obtain time-domain radial velocity (RV) measurements, we cross-matched the \textit{Gaia} astrometric candidates with the latest LAMOST DR13 v1.0 Medium-Resolution Survey (\citealt{lamostsurvey,lamostmrs}). To ensure the reliability of the spectroscopic parameters, we first excluded spectra with a signal-to-noise ratio ($\mathrm{S/N}$) below 10. A critical requirement for resolving long-period orbits is sufficient temporal sampling; thus, we discarded sources with fewer than five observational epochs to ensure a minimum level of temporal sampling. For the remaining sources, we performed an initial screening using the zero-point-corrected radial velocities (\texttt{rv\_r1}) provided by the LAMOST pipeline, filtering out systems with a peak-to-peak RV variation of less than $20~\mathrm{km~s^{-1}}$. This systematic refinement resulted in a spectroscopic sample of 4,516 sources.

For this sample, we developed a dedicated Python-based pipeline to extract precise relative RVs using the cross-correlation function (CCF) method. Each spectrum underwent simultaneous cleaning via a median filter and $3.5\sigma$ clipping to remove cosmic rays, followed by continuum normalization using anchor-point interpolation across 6400--6460~\AA\ and 6600--6660~\AA. Using the highest-$\mathrm{S/N}$ epoch as an internal template, we measured relative RVs through a multi-window CCF analysis centered on Fe~\textsc{i}, Ca~\textsc{i} (6380--6530~\AA), H$\alpha$ (6540--6590~\AA), and He~\textsc{i}/Li~\textsc{i} (6600--6750~\AA). The relative RV was determined as the median across these windows, with uncertainties estimated from the standard error and a minimum floor of $1.0~\mathrm{km~s^{-1}}$. 

To minimize the impact of stochastic errors, we implemented a $2.5\sigma$ median absolute deviation (MAD) filter to reject RV outliers. For the valid time series, the preliminary orbital period $P_{\mathrm{guess}}$ was searched using the Lomb-Scargle periodogram \citep{Lomb, scargle} within a range of 1.2 to 10,000 days, followed by a grid search to initialize non-linear least-squares fits. After the RV extraction and quality-control procedure, 1,057 of the 4,516 sources retained more than five valid RV measurements and were suitable for orbital fitting. Among these, 83 systems satisfied $f(M_2) > 0.05\,M_{\odot}$. We further screened these 83 systems by excluding double-lined binaries \citep{2025ApJS..278...46G} and systems without reliable orbital solutions. The latter included cases with periods reaching the search boundary, insufficient temporal coverage, poorly constrained orbital parameters, or large mass functions driven by individual RV outliers. This screening resulted in a final sample of 19 long-period single-lined binary candidates.

\subsection{SED Fitting}\label{sec:sed}

To characterize the physical properties of the visible primary stars, we constructed multi-wavelength SEDs for our sample. Photometric data were cross-matched using a coordinate-based search via the \texttt{astroquery.vizier} service \citep{vizier}. We searched the \textit{GALEX} GR6/7 catalog for far-ultraviolet (FUV) and near-ultraviolet (NUV) counterparts \citep{galex}. We retrieved optical magnitudes ($G$, $G_{\rm BP}$, $G_{\rm RP}$) from \textit{Gaia} DR3 \citep{2023A&A...674A...1G}, and near-to-mid-infrared magnitudes ($J$, $H$, $K_s$) from \textit{2MASS} \citep{2mass} and \textit{AllWISE} ($W1$, $W2$) \citep{allwise}. A search radius of $3''$ was applied for the \textit{Gaia}, \textit{2MASS}, and \textit{WISE} catalogs, while a wider radius of $5''$ was used for \textit{GALEX} to account for its larger positional uncertainty \citep{galex}. Although we systematically cross-matched our sample with the \textit{GALEX} catalog, none of the 19 candidates exhibited detectable far- or near-ultraviolet flux. Because the available \textit{GALEX} information corresponds to non-detections rather than measured fluxes, we do not include these bands in the SED fitting. The lack of UV counterparts is therefore treated as supporting, but not decisive, evidence against a hot luminous companion. Consequently, the subsequent SED fitting was performed on the eight optical-to-mid-infrared passbands (\textit{Gaia}, \textit{2MASS}, and \textit{WISE}). 

The theoretical photometric fluxes for these eight optical-to-mid-infrared passbands were synthesized using the MESA Isochrones and Stellar Tracks (MIST) models \citep{mist}. The parameter space was sampled using the affine-invariant Markov Chain Monte Carlo (MCMC) ensemble sampler \texttt{emcee} \citep{emcee}. For each candidate, we used 32 walkers and ran the chains for 100,000 steps, discarding the first 20,000 steps as burn-in. The resulting posterior samples were inspected through trace plots and used to estimate the stellar parameters and their uncertainties. The MCMC framework simultaneously sampled five free parameters: primary stellar mass $M_1$, age, metallicity $\mathrm{[Fe/H]}$, distance $d$, and $V$-band extinction $A_V$. To reduce the degeneracy between effective temperature and interstellar extinction, the joint likelihood function incorporated observational priors. Specifically, the LAMOST spectroscopic measurements $T_{\mathrm{eff}}, \log g$ and $\mathrm{[Fe/H]}$ were applied as Gaussian priors, while \textit{Gaia} DR3 parallaxes, corrected by adopting a $-0.025$\,mas zero-point offset, were used as priors on the geometric distance. This Bayesian analysis provided posterior estimates of the primary-star mass $M_1$ and bolometric luminosity $L_1$, which were then used to evaluate the expected luminosity of a putative main-sequence companion. To visually validate these fits, we matched the optimal stellar parameters to high-resolution BT-Settl synthetic spectra \citep{bt-settl}. The single-star models provide acceptable fits to the available optical-to-infrared photometry, with no significant broadband excess detected at the current photometric precision.

\subsection{Refined Orbital Solutions and Bayesian Dynamical Analysis} \label{sec:dynamics}

To refine the radial velocity measurements and place the RVs on an absolute velocity scale, we performed a template-matching procedure using the MARCS synthetic stellar atmosphere library \citep{marcs}. For each candidate, we identified the optimal synthetic template by minimizing the weighted distance in $T_{\mathrm{eff}}$--$\log g$--$\mathrm{[Fe/H]}$ space, using the best-fit parameters from our SED analysis as the primary constraints. These rest-frame synthetic spectra were employed as templates for a second-pass CCF analysis, yielding absolute RVs with a consistent and precise kinematic reference frame.

Based on these refined RVs, we employed a hybrid Bayesian framework to estimate orbital parameters. We first utilized \texttt{The Joker} \citep{thejoker}, a custom rejection sampler, to explore the multi-modal likelihood surface and identify candidate orbital solutions and possible multi-modal structures. These results served as initializations for a full posterior sampling using the native Hamiltonian Monte Carlo (HMC) No-U-Turn Sampler (NUTS) within \texttt{PyMC} \citep{pymc2023}. To optimize computational efficiency and avoid sampling pathologies near parameter boundaries, the eccentricity $e$ and argument of periastron $\omega$ were reparameterized using auxiliary variables $h = \sqrt{e}\cos\omega$ and $k = \sqrt{e}\sin\omega$. For each candidate, we executed four independent chains, with each chain consisting of 8,000 tuning steps followed by 4,000 sampling draws and a target acceptance probability of 0.95. The stellar-mass posterior $M_1$ derived from the SED fitting was subsequently combined with the mass-function posterior to infer the minimum companion mass. For each candidate in the final sample, we inspected the posterior samples, trace plots, and period distributions to identify possible multi-modal or alias-dominated solutions. Most candidates show well-behaved chains, while a small number of long-period systems exhibit mildly elevated MCMC diagnostic values despite single-peaked period posteriors. We therefore treat these orbital solutions as less tightly constrained, with their uncertainties reflected in the reported posterior intervals.

As an independent astrometric consistency check, we forward-modeled the expected \textit{Gaia} astrometric goodness of fit from the RV posterior samples. For each forward-model draw, the orbital inclination and longitude of the ascending node, which are unconstrained by the RV data, were sampled assuming an isotropic orientation, with $\cos i$ uniformly distributed over $[-1,1]$ and $\Omega$ uniformly distributed over $[0,2\pi)$. Under the dark-companion assumption, the photocentre motion was taken to follow the visible primary. We projected the resulting photocentre motion onto the \textit{Gaia} along-scan directions using the \textit{Gaia} scanning law, refitted a five-parameter single-star astrometric model, and converted the resulting goodness-of-fit distributions to the \textit{Gaia} RUWE scale for comparison with the \textit{Gaia} DR3 measurements.

\begin{figure}[ht!]
    \centering
    \includegraphics[width=0.85\columnwidth]{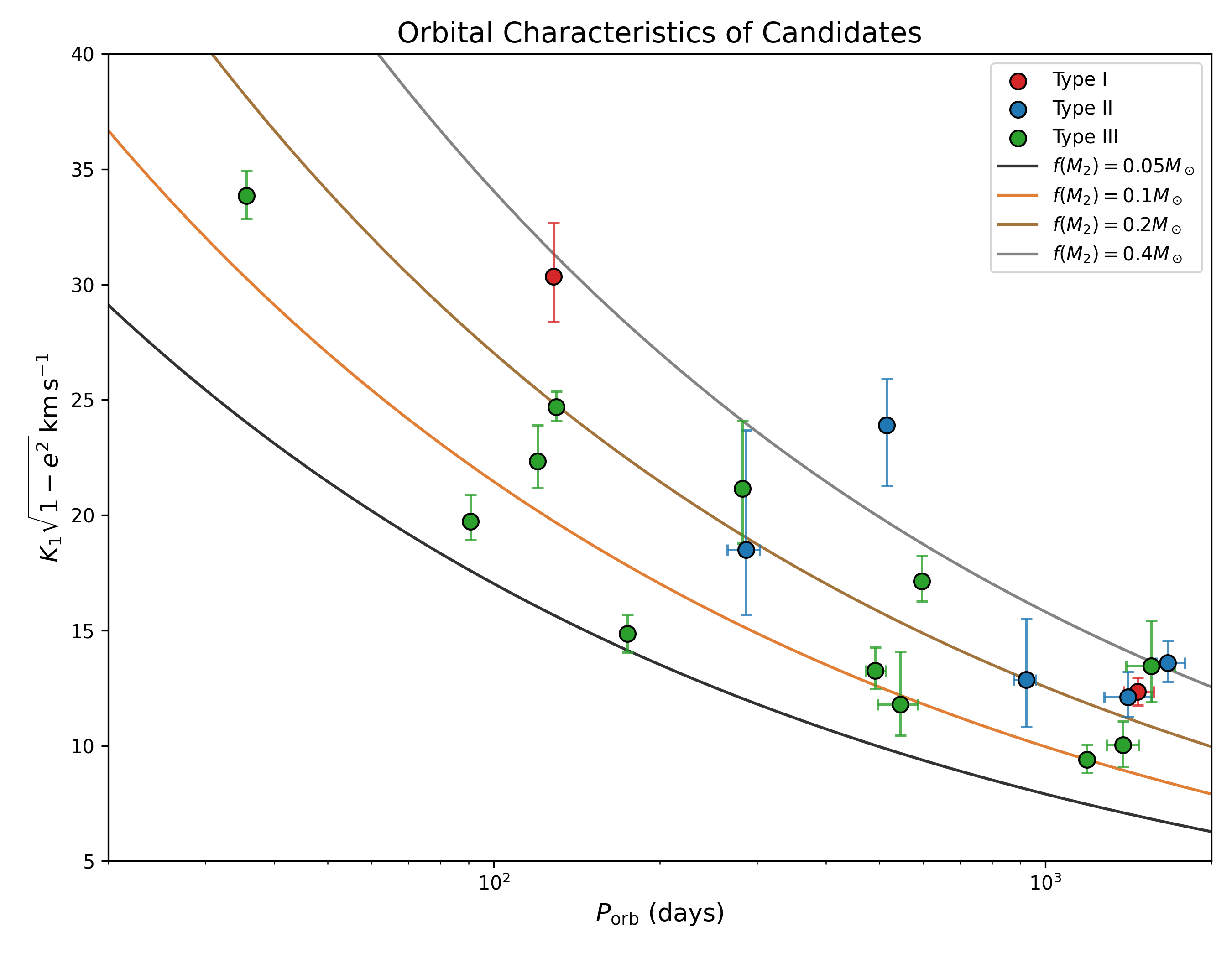}
    \caption{Distribution of the candidate systems in the $P_{\mathrm{orb}}$–$K_1\sqrt{1-e^2}$ plane. The solid lines represent constant mass functions $f(M_2)$ of 0.05, 0.1, 0.2, and 0.4 $M_\odot$, respectively. The candidates are color-coded by their classification based on the joint analysis: Type I (red), Type II (blue), and Type III (green). For some sources, the relative uncertainties in the orbital periods are sufficiently small that the corresponding horizontal error bars are not visually resolvable on the logarithmic scale at the current plotting resolution. All period uncertainties are included in the plot.}
    \label{fig:orbital_characteristics}
\end{figure}

For a single-lined spectroscopic binary, the binary mass function
provides a dynamical constraint on the mass of the unseen companion
\citep{fm1972PASP...84..292J, fm2001icbs.book.....H}. It can be expressed in terms of the observable orbital parameters as
\begin{equation}
    f(M_2)
    = \frac{M_2^3 \sin^3 i}{(M_1 + M_2)^2}
    = \frac{P_{\mathrm{orb}} K_1^3}{2\pi G}
      (1-e^2)^{3/2},
\end{equation}
where $M_1$ and $M_2$ are the masses of the visible primary and the
unseen companion, respectively, and $i$ is the orbital inclination.
For each posterior draw, we calculated the mass function using the
corresponding joint samples of the RV semi-amplitude $K_1$, orbital
period $P_{\mathrm{orb}}$, and eccentricity $e$. This procedure
naturally propagates both the individual parameter uncertainties and
their posterior correlations into the derived posterior distribution
of $f(M_2)$. Because the orbital inclination is generally unknown for a
single-lined spectroscopic binary, the mass function does not uniquely
determine the companion mass. Given an estimate of the primary-star
mass $M_1$, the minimum allowed companion mass, $M_{2,\min}$, is
obtained by assuming an edge-on configuration
($i=90^\circ$, $\sin i=1$). Any inclination smaller than
$90^\circ$ implies a larger companion mass. Figure~\ref{fig:orbital_characteristics} illustrates the distribution of our candidates in the $P_{\mathrm{orb}}$--$K_1\sqrt{1-e^2}$ plane. The solid lines denote constant mass functions ranging from $0.05$ to $0.4~M_\odot$. Based on the mass and luminosity-ratio constraints derived from the joint analysis, we categorized the systems into three groups: Type~I (2 sources, red circles) includes the two highest-priority compact-object candidates; Type~II (5 sources, blue circles) includes intermediate-priority systems requiring continued RV monitoring; and Type~III (12 sources, green circles) includes lower-priority systems for which faint stellar companions remain plausible.

\section{RESULTS AND DISCUSSION} \label{sec:results}

\subsection{Classification and Statistical Overview}
Through systematic joint astrometric-spectroscopic selection and comprehensive kinematic filtering, we identified a final sample of 19 long-period binary candidates. To assess the possible nature of the unseen companions, we utilized multi-wavelength SED fitting to characterize the physical parameters, notably the luminosity $L_1$ and mass $M_1$ of the visible primary stars. We then derived a theoretical luminosity ratio $L_2/L_1$ for each system under the assumption that the unseen secondary is a standard main-sequence star, calculating its expected luminosity $L_2$ based on the dynamical minimum mass $M_{2,\mathrm{min}}$ obtained from our MCMC orbital modeling. 

Restricting our focus to systems yielding significant mass functions ($f(M_2) > 0.05\,M_\odot$), and utilizing the derived mass ratios ($q = M_{2,\mathrm{min}}/M_1$) alongside these theoretical luminosity ratios, we classified the 19 sources into three distinct categories:

\begin{itemize}
    \item Type I (2 sources): These are the highest-priority compact-object candidates in our sample. Their nominal minimum mass ratios are $q=M_{2,\min}/M_1\simeq 1.08$ and $1.27$, indicating companions at least comparable in mass to the visible primaries. Under a main-sequence companion assumption, the corresponding estimated luminosity ratios are $L_2/L_1\simeq 2.19$ and $1.37$, respectively, implying that such companions should contribute detectable secondary light.
    \item Type II (5 sources): These systems form an intermediate-priority group: their dynamically inferred minimum companion masses are substantial, but their expected luminosity contributions under a main-sequence companion assumption are not high enough to rule out faint stellar companions. They exhibit minimum mass ratios ranging from $0.75$ to $1.17$, while their expected luminosity ratios are relatively low ($L_2/L_1<0.7$). This discrepancy makes them worthy of follow-up as compact-object candidates; however, faint stellar companions cannot yet be ruled out. Continued long-term RV monitoring is therefore required to refine their orbital solutions and clarify their nature.
    \item Type III (12 sources): These systems form the largest subgroup and are characterized by low expected luminosity ratios. Because the expected companion flux is low in the optical and infrared bands, additional multi-wavelength follow-up is required to determine whether the unseen companions are compact objects or faint stellar companions.
\end{itemize}

\subsection{Type I Candidates: Highest-Priority Compact-Object Candidates}
The two Type I systems, \textit{Gaia} DR3 3443894276188890752 (hereafter G3443) and \textit{Gaia} DR3 1560560462739675904 (hereafter G1560), are the highest-priority compact-object candidates in our catalog based on their relatively large minimum companion masses and the absence of detectable secondary light in the available data. The fundamental atmospheric and orbital parameters for the two Type I systems are presented in Table~\ref{tab:Type_i_parameters}. Our multi-wavelength SED fitting suggests that both visible primaries are consistent with G-type main-sequence stars, with inferred stellar masses of approximately $0.95\,M_\odot$ and $0.92\,M_\odot$, respectively. 

\begin{figure}[ht!]
    \centering
    \includegraphics[width=0.85\columnwidth]{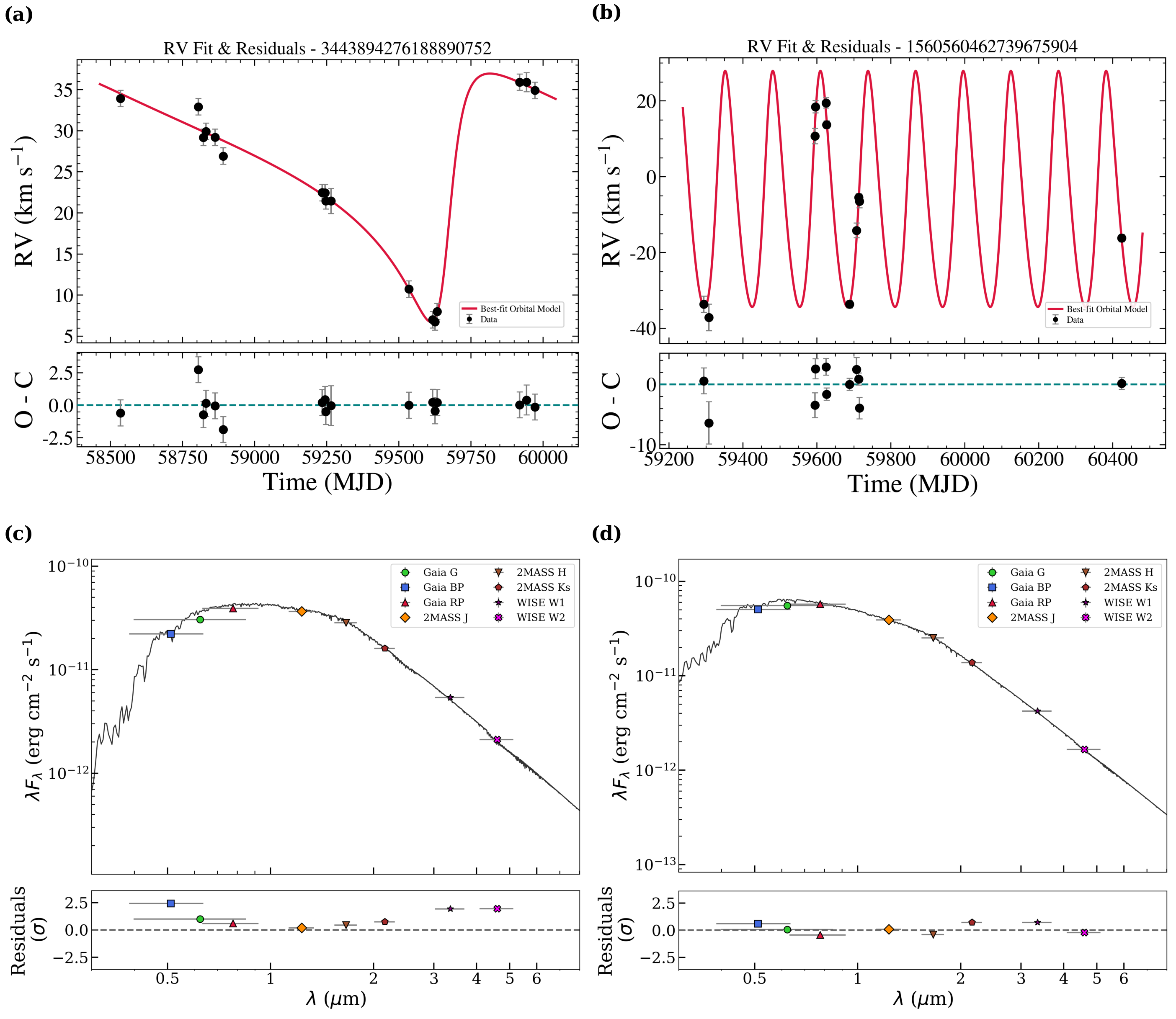}
    \caption{Dynamical and photometric modeling of the two Type I candidates. The upper panels show the radial-velocity observations and Keplerian orbital models. The lower panels display the corresponding multi-wavelength SED fits, combining photometry from optical to mid-infrared bands to constrain the primary stellar parameters ($M_1, R_1, T_{\mathrm{eff}}, \log g, [\mathrm{Fe/H}]$).}
    \label{fig:Type_i_rv_sed}
\end{figure}

The orbital dynamics and multi-wavelength photometric modeling for these two Type I systems are shown in Figure~\ref{fig:Type_i_rv_sed}. The upper panels display  Keplerian orbital solutions obtained with the hybrid \texttt{The Joker} and \texttt{PyMC} sampling framework.  G3443 has a longer period $P \approx 1469$ days, a higher eccentricity $e \approx 0.417$, and $f(M_2) \approx 0.285\,M_\odot$, yielding $M_{2,\mathrm{min}} \approx 1.04\,M_\odot$. G1560 has a period $P \approx 128$ days, an eccentricity $e \approx 0.137$, and mass function $f(M_2) \approx 0.371\,M_\odot$, corresponding to $M_{2,\mathrm{min}} \approx 1.18\,M_\odot$.

\begin{table}[ht!]
    \centering
    \caption{Orbital and Physical Parameters of Type I Candidates}
    \label{tab:Type_i_parameters}
    \begin{tabular}{lcc}
        \hline\hline
        Parameter & G3443 & G1560 \\
        \hline
        \textit{Gaia} DR3 Source ID  & 3443894276188890752 & 1560560462739675904\\
        LAMOST Designation   & J055403.02+300201.5 & J135042.43+520958.4 \\  
        RUWE  & 1.69 & 1.49\\ 
        $P_{\mathrm{orb}}$ (days) & $1469_{-81}^{+102}$  & $128_{-1}^{+1}$ \\
        $K_1$ (km s$^{-1}$) & $13.59_{-0.61}^{+0.62}$ & $30.64_{-1.97}^{+2.32}$ \\
        $e$     & $0.42_{-0.03}^{+0.04}$ &  $0.14_{-0.05}^{+0.06}$ \\
        $f(M_2)$ ($M_\odot$) & $0.285_{-0.041}^{+0.054}$ & $0.371_{-0.067}^{+0.090}$ \\
        $T_{\mathrm{eff}}$ (K)  & $5275_{-100}^{+99}$ & $5945_{-128}^{+132}$\\
        $\log g$    & $4.54_{-0.04}^{+0.02}$  & $4.30_{-0.04}^{+0.04}$\\
        $[\mathrm{Fe/H}]$ & $0.24_{-0.11}^{+0.10}$ & $-0.20_{-0.13}^{+0.14}$ \\
        $R_1$ ($R_\odot$) & $0.87_{-0.01}^{+0.02}$ & $1.12_{-0.02}^{+0.02}$ \\
        $L_1$ ($L_\odot$) & $0.53_{-0.04}^{+0.04}$ & $1.42_{-0.10}^{+0.11}$\\
        $M_1$ ($M_\odot$) & $0.95_{-0.07}^{+0.03}$ & $0.92_{-0.06}^{+0.08}$ \\
        $\varpi$ (mas) & $2.386_{-0.037}^{+0.037}$ & $1.526_{-0.017}^{+0.017}$ \\
        $M_{2,\mathrm{min}}$ ($M_\odot$) & $1.04_{-0.08}^{+0.10}$ & $1.18_{-0.13}^{+0.15}$ \\
        \hline\hline
    \end{tabular}
\end{table}

The lower panels of Figure~\ref{fig:Type_i_rv_sed} further show that the broadband photometry is consistent with a single-star stellar atmosphere model, with no significant infrared excess or anomalous optical-to-infrared flux that would indicate a luminous companion. For both systems, the derived minimum companion masses are compatible with massive white dwarfs and, depending on inclination, neutron stars. However, the current data do not uniquely determine the companion type, and faint stellar companions or unresolved hierarchical configurations cannot be fully excluded. Crucially, both systems remain single-lined in the LAMOST MRS spectra, with no double-peaked CCFs or secondary absorption features detected in the individual epochs. The absence of detected \textit{GALEX} counterparts and the lack of significant optical-to-infrared excess are consistent with this interpretation, but they do not by themselves confirm the compact-object nature of the companions.

\subsection{Nature and Follow-Up Requirements of Type II and Type III Sources}

The two Type I candidates are summarized in Table~\ref{tab:Type_i_parameters}, while the orbital and stellar parameters for the remaining 17 Type II and Type III candidates are provided in Table~\ref{tab:catalog}. The complete Table~\ref{tab:catalog} is available in machine-readable format.

\begin{table}[t]
\centering
\caption{Catalog of Type II and Type III Long-Period Binary Candidates for Compact-Object Follow-Up Observations}
\label{tab:catalog}
\scriptsize

\begin{tabular*}{\textwidth}{@{\extracolsep{\fill}}lcccccccc}
\tableline
\tableline
Gaia DR3 Source ID & RUWE & $P_{\rm orb}$ & $K_1$ & $f(M_2)$ & $M_1$ & $M_{2,\min}$ & $q$ & Type \\
                   &      & (days)        & (km s$^{-1}$) & ($M_{\odot}$) & ($M_{\odot}$) & ($M_{\odot}$) & & \\
\tableline
1893474166810285952 & 5.73 & 287  & 18.6 & 0.184 & 1.03 & 0.86 & 0.839 & II  \\
470096784872760192  & 3.13 & 923  & 13.1 & 0.203 & 1.39 & 1.08 & 0.779 & II  \\
3369415484530173696 & 1.93 & 1189 & 9.9  & 0.102 & 2.78 & 1.20 & 0.431 & III \\
2876854139510029184 & 2.56 & 546  & 11.8 & 0.090 & 1.48 & 0.76 & 0.517 & III \\
\tableline
\end{tabular*}

{\raggedright
\tablecomments{Only a portion of Table~\ref{tab:catalog} is shown here to demonstrate its form and content. The complete table, available in machine-readable format, contains 17 Type II and Type III candidates and includes coordinates, RUWE, parallaxes, stellar parameters, orbital solutions, mass functions, minimum companion masses, expected luminosity ratios, minimum mass ratios,} and follow-up priority types.
}
\end{table}

The five Type II systems occupy an intermediate parameter regime. Four of them are associated with evolved visible components, including G-/K-type subgiants and K-type giants, while one system has a G-type main-sequence primary. Their minimum companion masses range from $0.86$ to $2.42\,M_\odot$, with a mean value of $\sim 1.60\,M_\odot$. The corresponding minimum mass ratios span $q=0.75\text{--}1.17$, and the expected companion-to-primary luminosity ratios under the main-sequence assumption are $L_2/L_1=0.20\text{--}0.68$. These systems are therefore worthy of follow-up as compact-object candidates; however, faint stellar companions or hierarchical configurations cannot yet be ruled out.

The 12 Type III systems have a mean minimum companion mass of $M_{2,\min}\sim 1.18\,M_{\odot}$ and an average theoretical luminosity ratio of $\sim 0.073$. At such low luminosity ratios, the expected companion flux is small in the optical and infrared bands, making it difficult to distinguish compact objects from faint stellar companions using the current photometric data alone. These systems are therefore lower-priority but still astrophysically interesting follow-up targets. Determining the physical nature of their unseen companions will require additional observations, including ground-based high-resolution spectroscopy to search for spectral-line contamination and space-based ultraviolet imaging to test for hot compact companions.

We emphasize that this sample should not be interpreted as a complete census of compact-object binary candidates. The RUWE selection can be affected by unresolved multiplicity, crowding, and astrometric systematics, while the LAMOST multi-epoch requirement and RV-amplitude cut preferentially select systems with favorable temporal sampling and large projected velocity amplitudes. Our classification is therefore intended as a follow-up priority scheme rather than a definitive physical classification.

Using the astrometric forward-modeling procedure described in Section~\ref{sec:dynamics}, we compared the RV-predicted RUWE distributions with the Gaia DR3 measurements. Overall, 15 of the 19 systems are consistent with the Gaia measurements at the 90\% posterior-predictive level, including eight within the central 68\% interval, while four systems show larger discrepancies. For the two Type I candidates, the predicted RUWE values are $4.50^{+3.59}_{-2.04}$ for G3443 and $2.51^{+1.89}_{-0.45}$ for G1560, compared with the Gaia DR3 values of 1.69 and 1.49, respectively. G3443 also has a significant Gaia astrometric acceleration solution, whereas G1560 shows some quantitative tension between the RV-based prediction and the observed RUWE. These results indicate broad consistency between the RV-inferred orbital motion and the Gaia astrometric behavior, while a small number of systems show quantitative discrepancies.

\section{SUMMARY}
\label{sec:summary}

We have conducted a systematic search for long-period single-lined binaries that may host compact objects by combining \textit{Gaia} DR3 astrometric information with multi-epoch medium-resolution spectroscopy from LAMOST DR13. Starting from sources with elevated RUWE values and significant radial-velocity variability, we identified 19 single-lined long-period binary candidates and modeled their orbits using Bayesian Keplerian fits. We further constrained the properties of the visible stars through multi-wavelength SED fitting, allowing us to evaluate the dynamically inferred minimum mass and expected luminosity of the unseen companions under a main-sequence assumption.

Among the 19 systems, G3443 and G1560 are identified as the highest-priority compact-object candidates. G3443 has a wide and eccentric orbit, with $P_{\rm orb}=1469^{+102}_{-81}$ days, $e=0.42^{+0.04}_{-0.03}$, and $M_{2,\min}=1.04^{+0.10}_{-0.08}~M_\odot$. G1560 has a shorter-period orbit, with $P_{\rm orb}=128^{+1}_{-1}$ days, $e=0.14^{+0.06}_{-0.05}$, and $M_{2,\min}=1.18^{+0.15}_{-0.13}~M_\odot$. For both systems, the visible components are consistent with normal G-type main-sequence stars, while the expected flux from a main-sequence companion of the inferred minimum mass is not observed. Their SB1 nature, lack of significant optical-to-infrared excess, and absence of detected \textit{GALEX} counterparts make them high-priority compact-object candidates. Nevertheless, their physical nature remains unconfirmed, and follow-up spectroscopy, ultraviolet constraints, and future astrometric orbital solutions are required to distinguish massive white dwarfs, neutron stars, and faint stellar companions or hierarchical configurations.

The remaining candidates are less secure but still astrophysically interesting. Five Type II systems occupy an intermediate regime, with substantial minimum companion masses but theoretical companion-to-primary luminosity ratios of $0.20$--$0.68$ under the main-sequence assumption. Twelve Type III systems have much lower expected luminosity ratios, making their unseen companions compatible with either faint degenerate objects or stellar companions in less favorable orbital configurations. These 17 systems therefore require additional observations before their nature can be determined.

The availability of sufficiently numerous and high-quality multi-epoch LAMOST spectra represents a major observational bottleneck in our search, leaving only a small and potentially biased subset of the \textit{Gaia}-selected sources suitable for detailed RV analysis. Additional promising candidates may therefore remain outside the LAMOST footprint or below the effective sensitivity of the current spectroscopic data. Future high-resolution spectroscopy, continued radial-velocity monitoring, ultraviolet imaging, and forthcoming \textit{Gaia} astrometric orbital constraints will be essential for distinguishing compact objects from faint stellar companions. In particular, \textit{Gaia} DR4 may provide improved constraints on the orbital inclinations, thereby converting the RV-based minimum companion masses into more reliable true companion-mass estimates.

\begin{acknowledgments}

The authors thank Fabo Feng, Hao-Bin Liu, Qian-Yu An, Jiangxinxin Zhuang and Xianjin Shen for helpful suggestions, and thank the anonymous referee for the insightful suggestions. This work was supported by the National Key R\&D Program of China under grant 2023YFA1607901, and the National Natural Science Foundation of China under grants 12433007 and 12221003. This work has made use of data from the European Space Agency (ESA) mission {\it Gaia} (\url{https://www.cosmos.esa.int/gaia}), processed by the {\it Gaia} Data Processing and Analysis Consortium (DPAC, \url{https://www.cosmos.esa.int/web/gaia/dpac/consortium}). Funding for the DPAC has been provided by national institutions, in particular the institutions participating in the {\it Gaia} Multilateral Agreement. This work made use of the data from LAMOST (Large Sky Area Multi-Object Fiber Spectroscopic Telescope, also known as the Guoshoujing Telescope, \url{https://cstr.cn/31118.02.LAMOST}). LAMOST is a Chinese national mega-science facility, operated by National Astronomical Observatories, Chinese Academy of Sciences.
\end{acknowledgments}

\software{Astropy \citep{astropy:2013, astropy:2018, astropy:2022},  
          NumPy \citep{numpy}, 
          PyMC \citep{pymc2023},
          SciPy \citep{2020SciPy-NMeth},
          Spectool \citep{spectool},
          VizieR \citep{vizier},
          Matplotlib \citep{matplotlib},
          pandas \citep{the_pandas_development_team_2026_20127038}.
          }


\bibliography{sample701}{}
\bibliographystyle{aasjournalv7}



\end{document}